\documentclass[showpacs,aps,twocolumn]{revtex4}
\usepackage{amsmath}
\usepackage{times}
\usepackage{amssymb}
\usepackage{graphicx}
\usepackage{float}
\usepackage{color}
\def\<{\langle}
\def\>{\rangle}

\begin{document}
\today
\title{
Current fluctuations at a phase transition
}
 \author{Antoine Gerschenfeld and Bernard Derrida}
\affiliation{ Laboratoire de Physique Statistique, Ecole Normale
Sup\'erieure, UPMC Paris 6,Universit\'e Paris Diderot, CNRS,
24 rue Lhomond, 75231 Paris Cedex 05 - France}

\begin{abstract}
The ABC model is a simple diffusive one-dimensional non-equilibrium system which exhibits a phase
transition. Here we show that the cumulants of the currents of particles through the system become
singular near the phase transition. At the transition, they exhibit an anomalous dependence on the
system size (an anomalous Fourier's law). An effective theory for the dynamics of the single mode which
becomes unstable at the transition allows one to predict this anomalous scaling.
\end{abstract}


\maketitle

\date{\today}

A lot of work has been devoted recently to the study of the fluctuations of the current of heat or of
particles through non-equilibrium one dimensional systems
\cite{BD,BDGJL5,BBO,derrida2007,HS2,ADLW,ILW,HG,HG2,PM2,Tou,LM2}. In such studies the basic quantity one considers is   
the total flux $Q(t)$ of energy or of particles through a section of the system during time $t$.
In the steady state this flux $Q(t)$ fluctuates due to the randomness of the initial condition for
purely deterministic models and due to the noisy dynamics in stochastic models ({here we only
discuss} classical systems: see \cite{LLY,Be,BB,RB} for the quantum case).
If one assumes that the total energy or the total number of particles in the system remains bounded,
the average current $\lim_{t\to \infty} {\langle Q(t) \rangle \over t}$ as well as the higher cumulants
$\lim_{t\to \infty} {\langle Q(t)^n \rangle_c \over t}$ of the flux $Q(t)$ do not depend on the
section of the system {where this flux is measured}.

For a one dimensional system of length $L$, a central question is the size dependence of these
cumulants\cite{BLR}. In particular one would like to know whether a given system satisfies Fourier's 
law, meaning that{, for large $L$,} the average current scales like $1/L$:
\begin{equation}
\lim_{t\to \infty} {\langle Q(t) \rangle \over t} \simeq {A_1 \over L}
\label{fourier}
\end{equation}
where the prefactor $A_1$ depends on the temperatures $T_1$ and $T_2$ of the
two heat baths or on the chemical potentials $\mu_1$ and $\mu_2$ of the two reservoirs of particles
at the ends of the system.
At equilibrium ($T_1=T_2$ or $\mu_1=\mu_2$) the prefactor $A_1$ in (\ref{fourier}) vanishes { but}
the question of the validity of Fourier's law remains. One then wants to know whether the second
cumulant of $Q(t)$ scales like $1/L$.
\begin{equation}
\lim_{t\to \infty} {\langle Q(t)^2 \rangle - \langle Q(t) \rangle^2 \over t}=
\lim_{t\to \infty} {\langle Q(t)^2 \rangle_c \over t} \simeq {A_2 \over L}\,.
\label{fourier-bis}
\end{equation}
One can show that {(\ref{fourier-bis})} holds for diffusive systems such as the SSEP (symmetric simple
exclusion process)\cite{BD,derrida2007,DDR,HS} or the KMP (Kipnis-Marchioro-Presutti) model\cite{KMP}.
The macroscopic fluctuation theory developed by Bertini {\em et al.} \cite{BDGJL5,BDGJL6,BDGJL3} allows
{one} also to determine\cite{BD} all the cumulants of the flux $Q(t)$, with the result that they
all scale with system size as $1/L$.
\begin{equation}
\lim_{t\to \infty} {\langle Q(t)^n \rangle_c \over t} \simeq {A_n \over L}
\label{fourier-ter}
\end{equation}
Even corrections of order $1/L^2$ have been computed in some cases{\cite{ADLW,ILW}}.

For mechanical systems with deterministic dynamics, in particular systems which conserve momentum, the
average current scales as a non-integer power of the system size:
\begin{equation*}
\lim_{t\to \infty} {\langle Q(t) \rangle \over t} \simeq {B_1 \over L^{1- \alpha}}
\end{equation*}
The exponent $\alpha$ takes the value $1/2$ for some exactly soluble special models\cite{BBO}.{
Values ranging from 0.25 to 0.4 have also been reported} in simulations depending on the model
considered\cite{LW,MDN,GNY,Dhar2,GDL}. Theoretical { predictions} based on a mode coupling
approach\cite{LLP,LLP2} or on renormalization group calculations\cite{Dhar} confirm this anomalous
Fourier's law. Less is known on the size dependence of the higher cumulants, which are numerically
harder to measure, except that they vary as power laws of the system size, with exponents which
seem to depend on the geometry\cite{BDG}.

Here we consider the $ABC$ model\cite{EKKM,EKKM2}, a diffusive system which is known to exhibit a phase
transition\cite{CDE,ACLMMS,LM,BLS,LCM}: we study the fluctuations of the current near this transition.
Generically, outside the transition the cumulants have a diffusive scaling (\ref{fourier-ter}). Here,
we show that the amplitudes $A_n$ become singular as one approaches the transition, and that the
cumulants of $Q(t)$ exhibit anomalous scalings at the transition.
When the transition is second order, due to the destabilization of a single Fourier mode of the
density\cite{CDE}, the fluctuations in the whole critical regime can be understood in terms of a
Langevin equation for a single complex variable which represents the amplitude and the phase of this
Fourier mode.

\section{Definition}
The $ABC$ model\cite{EKKM,EKKM2,CDE,ACLMMS,LM,BLS,LCM} is a one-dimensional lattice gas, where each
site is occupied by one of three types of particles, $A$, $B$ and $C$. Neighboring sites exchange
particles at the rates
\begin{align*}
	AB & \underset{1}{\overset{q}{\rightleftharpoons}} BA\nonumber\\
	BC & \underset{1}{\overset{q}{\rightleftharpoons}} CB\\
	CA & \underset{1}{\overset{q}{\rightleftharpoons}} AC\nonumber
\end{align*}
with an asymmetry $q\leq 1$. Here, we consider the model on a ring of $L$ sites. Since the rates are
invariant under cyclic permutations of $\{A,B,C\}$ , most of the equations below will be written for a
species $a \in\{A,B,C\}$, with $b$ and $c$ denoting respectively the next and previous species. When
$q$ scales as
$$q=e^{-\beta/L}\,,$$
the { dynamics become diffusive}: { in particular, }the probability that site $1\leq k\leq L$ is occupied by a particle of type $a$ behaves as
\begin{equation*}
	{\rm Pro}[s_k(t) = a] \simeq \rho_a( k /L,t/L^2)\,,
\end{equation*}
where the macroscopic density profiles $\rho_a(x,\tau)$ follow a local Fourier's law : if $j_a(x,\tau)$
is the current associated to $\rho_a$, then
\begin{equation}\label{jdeter}
	j_a = -\partial_x\rho_a+\beta \rho_a(\rho_c-\rho_b)\,,
\end{equation}
which, together with the conservation law
\begin{equation}\label{conserv}
\partial_\tau \rho_a = -\partial_x j_a\,,	
\end{equation}
gives the hydrodynamic equations\cite{CDE} { satisfied by the density:}
\begin{equation}\label{hydrodeter}
	\partial_\tau \rho_a = \partial_x^2 \rho_a + \beta\partial_x \rho_a(\rho_b-\rho_c)\,.
\end{equation}
{ These equations conserve the fact that} $\sum_a \rho_a(x,\tau) = 1$ {(each site is occupied
by one of the three species)} and $\int_0^1 dx \rho_a(x,\tau) = r_a$, where $r_a$ is the total density
of particles of type $a$. The deterministic equations (\ref{jdeter}) and (\ref{hydrodeter}) are
only valid in the large $L$ limit, {for} diffusive time scales, i.e. $t\sim L^2$.

For small $\beta$, the constant density profiles $\rho_a(x) = r_a$ are a stable stationnary solution of
(\ref{hydrodeter}). { T}hese constant profiles become linearly unstable above a critical value
$\beta_*$\cite{CDE} given by
\begin{equation}\label{betac}
	\beta_* = {2\pi\over\sqrt{\Delta}} \;\;\mbox{ with }\;\; \Delta=1-2\sum_a r_a^2\,,
\end{equation}
so that the long-time limit of (\ref{hydrodeter}) {becomes a function $\bar\rho_a(x)$ of the space
variable $x$ for $\beta>\beta_*$}. It has been argued\cite{BD2} {(and checked numerically)} that,
in the steady state, these modulated profiles do not {move}.
If the phase transition to the modulated phase is second order, then it should
occur at $\beta=\beta_*$ { given by (\ref{betac})}. A first-order transition may { however}
occur at $\beta<\beta_*$ : this should { certainly\cite{CDE}} be the case at least for
$\Lambda<0$, with
\begin{equation}\label{Lambda}
	\Lambda = \sum r_a^2 - 2\sum r_a^3\,.
\end{equation}

In the following, we study the integrated current $Q_A(t)$ of $A$ particles through the 
system during time $t$:
\begin{align}\label{defqa}
	Q_A(t) &= {1\over L} \int_0^t dt' \int_0^1 dx\;j_A(x,t') dt' \nonumber\\
		   &= L\int_0^{t/L^2} d\tau \int_0^1 dx\; j_A(x,\tau) d\tau\,.
\end{align} 
This space average fluctuates with time, and we will be interested in its cumulants, as in
(\ref{fourier-ter}). Because the difference between the space average and the flux through a
section remains bounded, the cumulants of the flux through any section are the same as those of this
space average in the long time limit\cite{BD2}.

\begin{figure}[t]
  \includegraphics{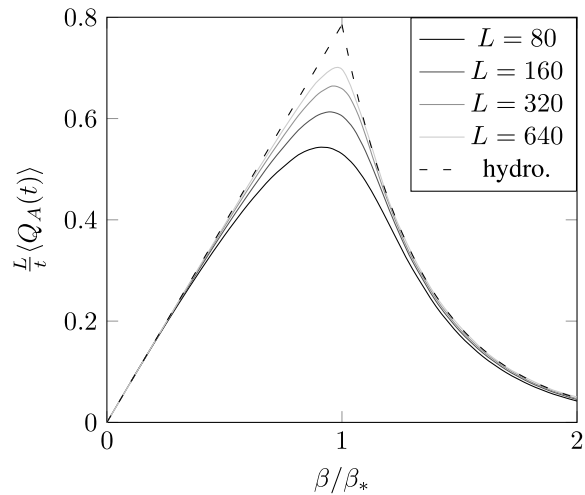}
  \caption{First cumulant of the integrated current of $A$ particles in the $ABC$ model with
  $r_A=r_B=1/4$, for $0\leq\beta\leq 2\beta_*$. Measurements of $\<Q_A\>$ in {Monte-Carlo
  simulations of} systems of $80$ to $640$ sites are compared to the prediction of the  hydrodynamic
  equation (\ref{qadeter}). }
\label{figc1}
\end{figure}

\section{Mean current}
As the steady-state profiles $\bar\rho_a(x)$ are time independent\cite{BD2},
(\ref{conserv}) implies that the steady-state currents are constant, $\bar j_a(x) = J_a$. They are
given by
\begin{equation}\label{qdeter}
	J_a = \beta\int_0^1 dx \bar\rho_a(x)(\bar\rho_c(x)-\bar\rho_b(x))\,.
\end{equation}
Then, from (\ref{defqa}), 
\begin{equation}\label{qadeter}
	{\<Q_A(t)\> \over t} \sim {{\beta}\over L} \int dx \bar\rho_A(x)(\bar\rho_C(x))-\bar\rho_B(x))\,.
\end{equation}
$\<Q_A(t)\>$ can thus be obtained by calculating numerically the long-time limit of (\ref{hydrodeter})
and then integrating (\ref{qdeter}). In Figure \ref{figc1}, we compare the results of this calculation
to numerical measurements obtained by simulating finite systems of $80$ to $640$ {sites}, for $r_A
= r_B = 1/4$ and $0\leq \beta \leq 2\beta_*${, with $\beta_*$ given by (\ref{betac})}.

For $\beta\leq\beta_*$, the stability of the flat density profiles $\bar\rho_a(x) = r_a$ leads to a
current $\<Q_A(t)\>\sim {t\over L}\beta r_A(r_C-r_B)$; on the other hand, the dependence in
$\beta$ becomes non-trivial for $\beta > \beta_*$, with a cusp at $\beta_*$.

For $\beta\downarrow\beta_*$, the steady-state profiles are known ({see} \cite{CDE} or (\ref{c1})
and (\ref{c1sc}) below) to take the form
\begin{equation}\label{sinus}
	\bar\rho_a(x) = r_a +\sqrt{\beta-\beta_*} (K_a e^{2i\pi x} + cc.) +{\mathcal O}(\beta-\beta_*)
\end{equation}
with known constants $K_A$, $K_B$ and $K_C$, leading to an analytic expression for $\<Q_A(t)\>$
around $\beta_*^+$:
\begin{equation*}
	{ L\over t}\<Q_A(t)\> \underset{\beta\downarrow\beta_*}{\simeq}  (r_C-r_B)\left[\beta r_A - 
	{\Delta^2\over\Lambda} (\beta-\beta_*)\right]\,. 
\end{equation*}

\begin{figure}[t]
  \includegraphics{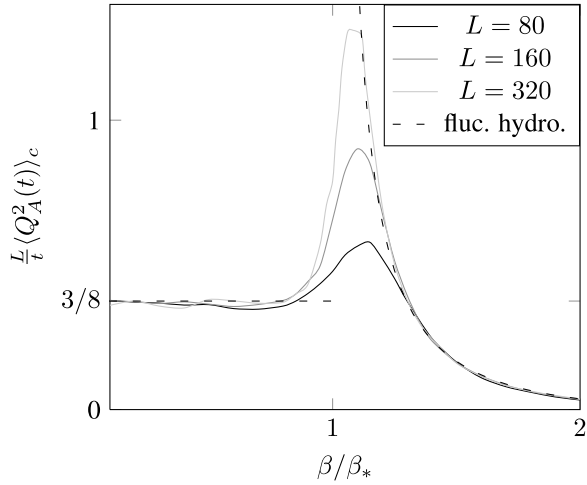}
  \caption{Second cumulant of the integrated current of $A$ particles in the $ABC$ model with
  $r_A=r_B=1/4$, for $0\leq\beta\leq 2\beta_*$. Measurements of $\<Q_A^2\>_c$ in systems
with $80$ to $320$ {sites} are compared to those obtained from the macroscopic fluctuation
theory (\ref{jfluc},\ref{qgen}). }
\label{figc2}
\end{figure}

\section{Fluctuation theory} The hydrodynamic equations (\ref{hydrodeter}) {describe the
deterministic evolution of the density profiles $\rho_a(x,\tau)$ in the large $L$ limit. F}or a large,
but finite system, {one has to take into account} stochastic corrections. {This can be done
using fluctuating hydrodynamics, where the expression (\ref{jdeter}) of the current is replaced
with}\cite{derrida2007,HS}
\begin{equation}\label{jfluc}
	j_a = q_a + {1\over\sqrt{L}}\eta_a(x,\tau)
\end{equation}
where $q_a = -\partial_x\rho_a + \beta\rho_a(\rho_c-\rho_b)$ is the right-hand side of (\ref{jdeter}),
and where the $\eta_a$ are Gaussian noises such that $\sum_a \eta_a = 0$ and
$$\<\eta_a(x,\tau)\eta_{a'}(x',\tau')\> = \sigma_{aa'}(x,\tau) \delta(x-x') \delta(\tau-\tau')\,,$$
with $\sigma_{aa} = 2\rho_a(1-\rho_a)$ and $\sigma_{aa'} = -2 \rho_a\rho_{a'}$ for $a\neq a'$. 
{Alternatively, the macroscopic fluctuation theory\cite{BDGJLX} expresses} 
(\ref{jfluc}) as a large deviation principle, {with} the probability of
observing {time-dependent} density profiles $\rho_a(x,\tau)$ given by
\begin{equation*}
	{\rm Pro}[\rho_a(x,\tau)] \propto \exp\left[ -L\iint dx d\tau\sum_a {(j_a - q_a)^2\over 4\rho_a}\right]\;.
\end{equation*}
From this formulation, the generating function of the current $Q_A(t)$ is the solution of the
optimization problem
\begin{equation}\label{qgen}
	\log\langle e^{\lambda Q_A(t)}\rangle = \max_{\rho_a,j_a} 
	L\int_0^{t/L^2}\hspace{-6mm}d\tau \int 
	dx\left[\lambda j_A-\sum_a {(j_a - q_a)^2\over 4\rho_a}\right]\,.
\end{equation}
Finding the density and current profiles $\rho_a(x,\tau)$ and $j_a(x,\tau)$ which maximize (\ref{qgen})
is not an easy task. We assume that, for large $t$, the optimum in (\ref{qgen}) is achieved by profiles
of fixed shape which may drift with a constant velocity $v$. In order to obtain $\<Q_A^2(t)\>_c$, we
compute $\log\langle e^{\lambda Q_A(t)}\rangle$ to order $2$ in $(\lambda,v)$, before optimizing over
$v$: one then gets optimization equations satisfied by these moving profiles, which we solved
numerically in the case $r_A=r_B=1/4$ for $0\leq\beta\leq 2\beta_*$.

For $\beta < \beta_*$, the constant profiles $\bar\rho_a(x) = r_a$ are still optimal for $\lambda\neq 
0$, yielding
$$\<Q_A^2(t)\>_c \simeq {2t\over L}r_A(1-r_A)\,. $$

For $\beta>\beta_*$, one has to take into account the dependence of the optimal profiles in $v$ and
$\lambda$. This leads (as will be shown in the longer \cite{DG2})
to a second cumulant which diverges at
$\beta=\beta_*$. This divergence can be computed exactly thanks to the knownledge of the steady-state
profiles (\ref{sinus}) for $\beta \downarrow \beta_*$, leading to
$$\< Q_A^2(t)\>_c \underset{\beta\downarrow\beta_*}{\simeq} {t\over L} {12\pi r_A r_B r_C (r_B-r_C)^2
\over \Lambda\sqrt{\Delta}(\beta-\beta_*)}\,.$$

\section{Critical regime for the deterministic hydrodynamics}
In this section, we analyse how the deterministic equations (\ref{hydrodeter}), exact {in the} $L\to\infty$ limit, behave in the neighborhood of $\beta_*$. The stability analysis of the constant
profiles $\rho_a = r_a$ shows that, as $\beta$ crosses $\beta_*$, only the first Fourier modes of
the $\rho_a$ become unstable{. For $\beta$ close to $\beta_*$, one therefore expects} this
mode {to} relax more slowly than the other Fourier modes.

One can write, from (\ref{hydrodeter}), an evolution equation for these slow
modes when $\beta$ is close to $\beta_*$, and show that they decay as a power law (instead of an
exponential) in the critical regime. To do so, we separate in $\rho_a(x,\tau)$ the
first Fourier mode, $R_a(\tau) e^{2i\pi x}$, from the other modes, $\tilde\rho_a(x,\tau)$, so that
\begin{align}
	\rho_a(x,\tau) &= r_a + R_a(\tau)e^{2i\pi x}+cc.+\tilde\rho_a(x,\tau)\,,\label{defrj}\\
	j_a(x,\tau) &= J_a(\tau) +{i\over2\pi}\dot R_a e^{2i\pi x}+cc. -\partial_\tau\int_0^x dy
	\tilde\rho_a(y,\tau)\nonumber
\end{align} 
with $\tilde\rho_a \ll R_a \ll r_a$. We also suppose that $\rho_a(x,\tau)$ varies slowly, so
that  $\partial_\tau R_a\equiv \dot R_a \ll R_a$ and $\partial_\tau\tilde\rho_a \ll \tilde\rho_a$; 
finally, we set
\begin{equation*}
	\beta = \beta_*(1+\gamma)
\end{equation*}
with $\gamma\ll 1$. The leading order of (\ref{jdeter}) then becomes, when projected on the
first and second Fourier modes,
\begin{align}
	2i\pi R_a =& \beta_*[R_a(r_c-r_b) + r_a(R_c-R_b)]\label{1per1}\\
	\partial_x\tilde\rho_a =& \beta_*[\tilde\rho_a(r_c-r_b)+ 
	r_a(\tilde\rho_c-\tilde\rho_b)\nonumber\\
	&+R_a(R_c-R_b)e^{4i\pi x}+cc.]\label{2per1}
\end{align}
Equation (\ref{1per1}) relates the leading orders of $R_B$ and $R_C$ to $R_A$, so that
\begin{equation}\label{Ri}
	\left\{ \begin{array}{l}
		R_B(\tau) = {2r_C-1-i\sqrt{\Delta}\over2r_A}R_A(\tau)+x_B(\tau)\\
		R_C(\tau) = {2r_B-1+i\sqrt{\Delta}\over2r_A}R_A(\tau)+x_C(\tau)
	\end{array}\right.
\end{equation}
with $x_B, x_C\ll R_A$.
Equation (\ref{2per1}) shows that, at {leading} order, $\tilde\rho_a$ is of the form
\begin{equation}\label{defphi}
	\tilde\rho_a(x,\tau) = \varphi_a e^{4i\pi x} + cc.
\end{equation}
with
$$4i\pi\varphi_a = \beta_*[\varphi_a(r_c-r_b)+r_a(\varphi_c-\varphi_b)+R_a(R_c-R_b)]\,,$$
{whose solution is}
\begin{equation}\label{phi}
	\varphi_a = {1-2r_a\over \Delta} R_a^2\,.
\end{equation}
Then, the next-to-leading order of the first Fourier mode of (\ref{jdeter}) reads
\begin{align}
	{i\over 2\pi} \dot R_a =&2i\pi\gamma R_a+\beta_*[(r_c-r_b-i\sqrt{\Delta})x_a + 
	r_a(x_c-x_b)\nonumber\\
	&+\varphi_a(R^*_c 
	-R^*_b)+R_a^*(\varphi_c-\varphi_b)]\label{1per3}
\end{align}
(with $x_A\equiv 0$).
The $x_a$ can be eliminated by multiplying the equation over $\dot R_A$ by $(r_C-r_B
+i\sqrt{\Delta})$, the one over $\dot R_C$ by $r_A$, the one over $\dot R_B$ by $-r_A$, and by
summing, which yields
\begin{equation}\label{evoldeter}
	\dot R_A = 4\pi^2\left(\gamma-{2\Lambda\over\Delta^2}{|R_A|^2\over r_A}\right) R_A\,,
\end{equation}
with $\Lambda$ {given} by (\ref{Lambda}).

For $\beta=\beta_*$ ($\gamma=0$), the first Fourier mode $R_A$ decays as a power law instead of an
exponential:

$$R_A(\tau) = {R_A(0)\over \sqrt{1+{16\pi^2\Lambda\over r_A \Delta^2}|R_A(0)|^2 \tau}}\,, $$
with an amplitude which does not depend on the initial condition for large $\tau$.

\section{Critical behavior of the MFT}
We now return to the noisy equation (\ref{jfluc}) and try to obtain a noisy version of
(\ref{evoldeter}). By analogy with the deterministic case above, we suppose 
that the first Fourier mode $R_a$ of $\rho_a$ varies more slowly, but with a larger amplitude
than the other modes,  so that $\dot R_a \ll R_a$ and $\tilde \rho_a \ll R_a$ in (\ref{defrj}).
When replacing (\ref{jdeter}) with (\ref{jfluc}) ,the leading-order equation (\ref{1per1}) is not
modified, so that (\ref{Ri}) still holds : however, the next-to-leading order equation (\ref{1per3})
is replaced with
\begin{align}
	{i\over 2\pi} \dot R_a =&2i\pi\gamma R_a+\beta_*[ 
	(r_c-r_b-i\sqrt{\Delta})x_a + r_a(x_c-x_b) \nonumber\\
	&+\varphi_a(R^*_c-R^*_b)+R_a^*(\varphi_c-\varphi_b)]+ {\nu_a(\tau)\over \sqrt{L}}\,,\label{1per3b}
\end{align}
where the $\nu_a$ are projections of the $\eta_a(x,\tau)$ on the first Fourier mode:
$$\nu_a(\tau) = \int_0^1 dx e^{-2i\pi x}\eta_a(x,\tau)\,, $$
so that $\<\nu_a(\tau)\nu^*_{a'}(\tau')\> = \sigma_{aa'}\delta(\tau-\tau')$ and
$\< \nu_a(\tau) \nu_{a'}(\tau')\>=0$. The second Fourier modes $\varphi_a$ (\ref{defphi}) satisfy the
equations
\begin{align*}
	{i\over 4\pi}\dot\varphi_a =& -4i\pi\varphi_a +\beta_*[\varphi_a(r_c-r_b)+r_a(\varphi_c-\varphi_b)\\
	&+R_a(R_c-R_b)] + {\nu_a^{(2)}(\tau)\over\sqrt{L}}\,,
\end{align*}
where the $\nu_a^{(2)}$ are projections of the $\eta_a(x,\tau)$ as well : hence, they fluctuate around
their non-noisy expression (\ref{phi}). In (\ref{1per3b}), however, these fluctuations (of amplitude
$1/\sqrt{L}$) are multiplied by $R_a^*$, so that they are of smaller amplitude than the noisy term
$\nu_a(\tau)/\sqrt{L}$ : therefore, the $\varphi_a$ can be replaced by their expression (\ref{phi}) in
(\ref{1per3b}).

Taking a linear combination of (\ref{1per3b}) to eliminate the $x_a$ as in the deterministic case
(\ref{1per3}), we then obtain
$$\dot R_A = 4\pi^2\left(\gamma-{2\Lambda\over\Delta^2}{|R_A|^2\over r_A}\right) R_A + {\mu_A(\tau)\over \sqrt{L}}\,,$$
with $\mu_A$ a linear combination of the $\nu_a$, which verifies
$$\<\mu_A(\tau)\mu_A^*(\tau')\> = {24\pi^2 r_A^2r_Br_C\over\Delta}\delta(\tau-
\tau')\,. $$
Finally, the change of variables
\begin{multline}\label{rescale}
	R_A(\tau) = \sqrt[4]{3\Delta r_A^3 r_B r_C\over \Lambda L} f(\bar\tau) \\\mbox{ with }
	\bar\tau=8\pi^2{\sqrt{3\Lambda r_A r_B r_C}\over \Delta^{3/2}}{\tau\over\sqrt{L}}
\end{multline}

leads to a simple rescaled equation:
\begin{multline}\label{critfluc}
	{df\over d\bar\tau} = (\bar\gamma-|f(\bar\tau)|^2)f(\bar\tau) + \mu(\bar\tau)\\
	\mbox{with } \bar\gamma = \sqrt{L}{\Delta^{3/2}\over 2\sqrt{3\Lambda r_A r_B r_C}}{\beta-\beta_*
	\over\beta_*}
\end{multline}
and with $\mu$ such that $\<\mu(\bar\tau)\mu^*(\bar\tau')\> = \delta(\bar\tau-\bar\tau')$.

Therefore, a system of size $L$ exhibits a critical regime $|\beta-\beta_*|\sim 1/\sqrt{L}$ in which
the density profiles $\rho_a(x,\tau)$ fluctuate as sine waves of period $1$, with an amplitude scaling
as $1/L^{1/4}$ on a time scale of order $1/\sqrt{L}$. The rescaled fluctuations, 
$f(\bar\tau)$, follow a (complex) damped Langevin dynamics in the quartic potential
$$ V(f) = -\bar\gamma{|f|^2\over 2} + {|f|^4\over 4}\,,$$
and the probability distribution of $f(\bar\tau)$, $P(r,\theta,\bar\tau) = {\rm Pro}[f(\bar\tau) 
\simeq r e^{i\theta}]$, satisfies the Fokker-Planck equation
\begin{equation}\label{fkp}
	\partial_{\bar\tau}P = {1\over r}\partial_r\left[(r^2-\bar\gamma)r^2P + {1\over4}r\partial_rP\right] + {1\over 4r^2}\partial_\theta^2 P\,.
\end{equation}

\begin{figure}[t]
  \includegraphics{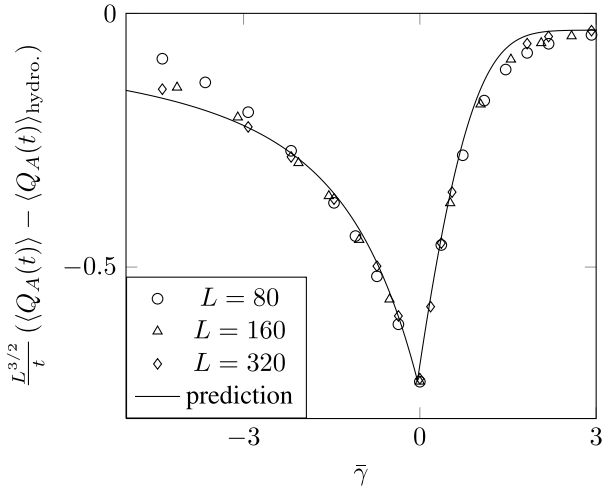}
  \caption{Rescaled deviation of the first cumulant of $A$ particles, $\<Q_A(t)\>$, from its
deterministic hydrodynamics prediction (\ref{qadeter}), as a function of $\bar\gamma$
(\ref{critfluc}), for $r_A=r_B=1/4$. Numerical measurements for systems of $80$ to $320$ sites
are compared with our analysis of the critical point (\ref{c1},\ref{c1sc}).  }
\label{figc1sc}
\end{figure}

\section{Critical fluctuations of the current}
Let us now discuss the consequences of the slow fluctuations of the first Fourier mode of the density
described above on the integrated particle current of $a$ particles, $Q_a(t)$.
From (\ref{jfluc}) and (\ref{defrj}), we can express the average instantaneous current, $J_a$, in terms
of $f(\bar\tau)$ :
\begin{multline*}
	J_a(\tau) = \beta r_a(r_b-r_c) + 2\beta(r_b-r_c)\sqrt{3\Delta r_ar_br_c\over \Lambda L}|f(\bar\tau)|^2\\
	+{1\over \sqrt{L}} G_a(\tau) + {\mathcal O}\left(1\over L\right)
\end{multline*}
where $G_a$, the space average of the noise $\eta_a$, is such that
$$\<G_a(\tau) G_a(\tau')\> = 2 r_a(1-r_a){\delta(\tau-\tau')} + {\mathcal O}(1/\sqrt{L})\,.$$
Therefore, the contributions of the fluctuations of the first Fourier mode $f$ and of the noise $G_a$
to $J_a$ are of comparable amplitude. The fluctuations of $f$, however, occur on the slower time scale
$\bar\tau\sim \tau/\sqrt{L}$: hence, they become dominant in the integrated current (\ref{defqa})
$$Q_a(t) = L\int_0^{t/L^{2}} J_a(\tau)d\tau\,, $$
with the $n$-points time correlation function of $f(\bar\tau)$ giving rise to an anormal growth of the
$n$-th cumulant of $Q_a$, $\<Q_a^n(t)\>_c$. More precisely, we find that
\begin{equation}\label{c1}
	\<Q_a(t)\> \simeq {t\over L} \beta r_a(r_c-r_b) + {2t\over L^{3/2}}\beta (r_b-r_c)\sqrt{3\Delta 
	r_ar_br_c
	\over \Lambda} C_1(\bar\gamma)
\end{equation}
and
\begin{equation*}
	\<Q_a^n(t)\>_c \simeq {t\over L^{5/2-n}} {8\pi^2\sqrt{3\Lambda r_a r_b r_c}\over \Delta^{3/2}}\left[
	\Delta^{3/2}(r_b-r_c)\over 2\pi\Lambda\right]^n\hspace{-2.5mm} C_n(\bar\gamma)
\end{equation*}
with $\bar\gamma$ as defined in (\ref{critfluc}) and with 
\begin{equation*}
	C_n(\bar\gamma) = \lim_{\bar\tau\to\infty} {1\over\bar\tau} \int_0^{\bar\tau}
	d\bar\tau_1 .. d\bar\tau_n \<|f(\bar\tau_1)..f(\bar\tau_n)|^2\>_c
\end{equation*}  
a time integral of the $n$-point correlation function of $f(\bar\tau)$.

Because $f(\bar\tau)$ follows a Langevin equation in a quartic potential (\ref{critfluc}), one does
not have simple analytical expressions for the $C_n(\bar\gamma)$ for $n\geq 2$. $C_1(\bar\gamma)$, on the other hand,
only depends on the stationary average of $|f(\bar\tau)|^2$: since the stationnary state of the
Fokker-Planck equation (\ref{fkp}) is $P(r,\theta)\propto e^{2\bar\gamma r^2-r^4}$, we obtain easily
that
\begin{equation}\label{c1sc}
	C_1(\bar\gamma) = \bar\gamma + {1\over 2} {e^{-\bar\gamma^2}\over\int_{-\infty}^{\bar\gamma} e^{-z^2} 
	dz}\,.
\end{equation}

Thus, we obtain an analytic expression for the deviation of $\< Q_A(t) \>$ from its hydrodynamics
prediction (\ref{qadeter}) in the neighborhood of $\beta_*$. In Figure \ref{figc1sc}, we 
compare this expression to numerical simulations of systems of $80$ to $320$ sites, for $r_A = r_B = 1/4$.

\section{Conclusion}
In this letter we have seen that a driven diffusive system like the ABC model, at a second order phase
transition, may exhibit anomalous Fourier's law at least for the second and higher cumulants.
This is reminiscent of the cumulants of the current which diverge with the system size in the TASEP
(the totally asymmetric exclusion process) along the first order transition line\cite{LM2}.
The mechanism is however different. Here the large fluctuations can be understood by analysing the
dynamics of the first Fourier mode which becomes unstable at the transition whereas, in the TASEP, the
large fluctuations of the current are due to the presence of the shock.

The anomalous current fluctuations of the ABC model at the phase transition are accompanied by
anomalous long range density fluctuations which can also be understood in terms of the slow noisy
dynamics of the first Fourier mode (the density fluctuations will be discussed in the forthcoming
longer version of the present letter\cite{DG2}).

An interesting open question would be to compare the anomalous density and the current fluctuations of
the ABC model at the transition with those of momentum conserving mechanical models, in particular
through the dynamics of their slow modes. Another interesting question would be to study the current
fluctuations through other lattice gases (such as an Ising model) when there is coexistence of several
phases at equilibrium.

\end{document}